\documentclass[11pt,a4paper]{article} 
\pdfoutput=1
 \usepackage{jheppub}
\allowdisplaybreaks

\begin{document}

\preprint{ULB-TH/15-26}

\title{Quark Seesaw, Vectorlike Fermions and Diphoton Excess}
\author[a]{P. S. Bhupal Dev,}
\author[b]{Rabindra N. Mohapatra,}
\author[c]{and Yongchao Zhang}

\affiliation[a]{ Max-Planck-Institut f\"{u}r Kernphysik, Saupfercheckweg 1, D-69117 Heidelberg, Germany}
\affiliation[b]{Maryland Center for Fundamental Physics, Department of Physics, University of Maryland, College Park, MD 20742, USA}
\affiliation[c]{Service de Physique Th\'{e}orique, Universit\'{e} Libre de Bruxelles,
Boulevard du Triomphe, CP225, 1050 Brussels, Belgium}

\emailAdd{bhupal.dev@mpi-hd.mpg.de}
\emailAdd{rmohapat@umd.edu}
\emailAdd{yongchao.zhang@ulb.ac.be}

\date{\today}

\abstract{
  We present a possible interpretation of the recent diphoton excess reported by the early $\sqrt s=13$ TeV LHC data in  quark seesaw left-right models with vectorlike fermions proposed to solve the strong $C\!P$ problem without the axion. The gauge singlet real scalar field responsible for the mass of the vectorlike fermions has the right production cross section and diphoton branching ratio to be identifiable with the reported excess at around 750 GeV diphoton invariant mass. Various ways to test this hypothesis as more data accumulates at the LHC are proposed. 
}

\maketitle


\section{Introduction}
In the early run II data from the $\sqrt s=13$ TeV Large Hadron Collider (LHC), both CMS~\cite{CMS} and ATLAS~\cite{ATLAS} experiments have reported an excess of $\gamma\gamma$ events over the SM background with invariant mass around 750 GeV. The signal cross section is reported to be $(6\pm 3)~{\rm fb}$ by CMS~\cite{CMS} and $(10\pm 3)~{\rm fb}$ by ATLAS~\cite{ATLAS}. While this excess has a local statistical significance of around $2.6\sigma$ (CMS) to 3.9$\sigma$ (ATLAS) and needs more data to firmly rule out the possibility of a statistical fluctuation, it has nonetheless generated a great deal of recent activity in the theory community as a possible signal of beyond the Standard Model (SM) physics and many possible interpretations have been advanced; for a non-exhaustive list of ideas and speculations, see~\cite{Harigaya:2015ezk,
Mambrini:2015wyu,
Backovic:2015fnp,
Angelescu:2015uiz,
Nakai:2015ptz,
Knapen:2015dap,
Buttazzo:2015txu,
Pilaftsis:2015ycr,
Franceschini:2015kwy,
DiChiara:2015vdm,
Higaki:2015jag,
Molinaro:2015cwg,
Petersson:2015mkr,
Gupta:2015zzs,
Bellazzini:2015nxw,
Low:2015qep,
Ellis:2015oso,
Agrawal:2015dbf,
Cox:2015ckc,
Ahmed:2015uqt,
Kobakhidze:2015ldh,
Matsuzaki:2015che,
Cao:2015pto,
Dutta:2015wqh,
Bai:2015nbs,
Aloni:2015mxa,
Falkowski:2015swt,
Csaki:2015vek,
Chakrabortty:2015hff,
Bian:2015kjt,
Curtin:2015jcv,
Fichet:2015vvy,
Chao:2015ttq,
Demidov:2015zqn,
No:2015bsn,
Becirevic:2015fmu,
Martinez:2015kmn,
Carpenter:2015ucu,
Bernon:2015abk,
Megias:2015ory,
Alves:2015jgx,
Kim:2015ron,
Benbrik:2015fyz,
Gabrielli:2015dhk,
Chao:2015nsm,
Arun:2015ubr,
Han:2015cty,
Chang:2015bzc,
Chakraborty:2015jvs,
Ding:2015rxx,
Han:2015dlp,
Han:2015qqj,
Luo:2015yio,
Chang:2015sdy,
Bardhan:2015hcr,
Feng:2015wil,
Antipin:2015kgh,
Wang:2015kuj,
Cao:2015twy,
Huang:2015evq,
Liao:2015tow,
Heckman:2015kqk,
Bi:2015uqd,
Kim:2015ksf,
Berthier:2015vbb,
Cho:2015nxy,
Cline:2015msi,
Chala:2015cev,
Barducci:2015gtd,
Bauer:2015boy,deBlas:2015hlv,
Dev:2015isx, Boucenna:2015pav, McDermott:2015sck, Murphy:2015kag, Hernandez:2015ywg, Dey:2015bur, Pelaggi:2015knk, Cao:2015xjz, Huang:2015rkj, Moretti:2015pbj, Patel:2015ulo, Badziak:2015zez, Chakraborty:2015gyj, Cvetic:2015vit, Allanach:2015ixl, Das:2015enc, Cheung:2015cug, Davoudiasl:2015cuo, Liu:2015yec, Zhang:2015uuo, Casas:2015blx, Hall:2015xds, Altmannshofer:2015xfo, Gu:2015lxj, Craig:2015lra}. In this note we add another one in the context of a theory proposed many years ago as a solution to the strong $C\!P$ problem without an axion~\cite{BM, BM1}.

The model is based on the assumption that there exist TeV-scale vectorlike fermions which are responsible for the seesaw masses for the quarks and charged leptons~\cite{univ1, univ2, univ3, univ4} in the context of a left-right (LR)  symmetric model based on the gauge group $SU(3)_C \times SU(2)_L\times SU(2)_R\times U(1)_{B-L}$~\cite{LR1, LR2, LR3}. The SM fermions and the vectorlike fermions are assigned to the gauge group as follows:
 \begin{eqnarray}
{\rm SM~ fermions}:&& Q_{L,R} \ = \ \left(\begin{array}{c} u\\ d\end{array}\right)_{L,R}, \quad \psi_{L,R} \ = \ \left(\begin{array}{c} \nu\\ e\end{array}\right)_{L,R} ; \nonumber \\
 {\rm Vectorlike~ fermions}:&& P\left({\bf 3}, {\bf 1},{\bf 1},+\frac{4}{3}\right), \quad  N\left({\bf 3}, {\bf 1}, {\bf 1}, -\frac{2}{3}\right), \quad  E({\bf 1},{\bf 1},{\bf 1},-2).
 \end{eqnarray}
The Higgs sector of the model consists of $SU(2)_{L,R}$ doublets $\chi_{L,R}$  which break the left and right $SU(2)$'s and a real singlet $S$ that gives mass to the vectorlike fermions. An appropriate discrete $Z_2$ symmetry forbids the bare mass of the vectorlike fermions. Under this $Z_2$ symmetry, the Higgs fields $\chi_L$ and $S$ are odd as are the right-handed (RH) chirality of the vectorlike  fermions; all other fields are even. The Yukawa couplings are given in this case by the Lagrangian
\begin{eqnarray}
\label{eq:Lyukawa}
- \mathcal{L}_Y & \ = \ &
 y_U  \bar{Q}_L \tilde{\chi}_L P_R
+ y_D\bar{Q}_L  \chi_L N_R
+ y_E \bar{L}  \chi_L E_R
+ (L \leftrightarrow R) \nonumber \\
&&+ f_U \bar{P}_L S P_R
+ f_D \bar{N}_L S N_R
+ f_E \bar{E}_L S E_R
+ {\rm H.c.} \,.
\end{eqnarray}
where $\tilde{\chi}_{L,R} = i\sigma_2\chi^*_{L,R}$ ($\sigma_2$ being the second Pauli matrix),
and $y_{F}$, $f_{F}$ (with $F=U,D,E$) are the Yukawa couplings with potential beyond SM $C\!P$  violations. Once both the doublets and the singlet obtain their non-vanishing vacuum expectation values (VEVs) $v_L,~v_R,~v_S$ respectively, we get the seesaw form for the $2\times2$ mass matrix for a single quark or lepton flavor:
\begin{eqnarray}
{\cal M}_F \ = \ \left( \begin{matrix}
0 & \frac{1}{\sqrt2} y_F v_L \\ \frac{1}{\sqrt2} y_F v_R & f_F v_S
\end{matrix} \right) \,,
\label{mass}
\end{eqnarray}
which generates the small fermion masses in the SM (except the top-quark mass) and alleviates the hierarchy problem in the Yukawa couplings:
\begin{eqnarray}
\label{seesaw}
m_{F} \ \simeq \ \frac{y_F^2 v_L v_R}{2 f_F v_S} \,.
\end{eqnarray}
Both the two new VEVs are assumed to be at the (multi-)TeV scale, 
whereas $v_L \simeq 246.2~{\rm GeV}$ is the electroweak scale. 
Clearly, the simple seesaw mass  formula in Eq.~(\ref{seesaw}) is not a good approximation for the top quark, as it is expected that the matrix entries $y_F v_R$ and  $f_F v_S$ are of similar magnitude, which implies a large ``right-handed'' mixing of the top quark and its partner through $\sin\alpha_R^t \sim \frac{1}{\sqrt2} y_T v_R / f_T v_S$. Therefore in general, one should take into consideration the whole $2\times2$ mass matrix \eqref{mass} and diagonalize ${\cal M}_F{\cal M}_F^\dagger$ to get the mass eigenvalues of the SM quarks and their partners. 

As far as the flavor structure and quark mixing are concerned, we can have either heavy quarks with degenerate masses (respectively for the up and down type flavors) in which case the SM quark mixings are completely determined by the flavor structure of the Yukawa couplings $y_{U,\,D}$ \cite{Mohapatra:2014qva}, or the couplings $y_{U,\,D}$ are hierarchical but diagonal (e.g. from some discrete symmetry assignments) and the matrices $f_{U,\,D}$ are of order ${\cal O}(1)$ in which case we have flavor anarchic \cite{D'Agnolo:2015uta}.

As a direct result of the Lagrangian in (\ref{eq:Lyukawa}), the heavy vectorlike quarks decay dominantly to the SM gauge and Higgs bosons plus SM quarks, especially for the top and bottom partners. Due to the Goldstone equivalence theorem, the branching ratios for the decays to $W$, $Z$ and Higgs are approximately $2:1:1$. The current LHC constraints put a 95\% confidence level (CL) lower limit on the top partner mass from 715--950 GeV and on the bottom partner mass from 575--813 GeV~\cite{Aad:2015kqa}, depending on their decay branching ratios. These lower bounds are much stronger than those from the flavor changing neutral currents (FCNCs) mediated by the heavy quarks~\cite{Mohapatra:2014qva}.

\section{Scalar sector}
The gauge and $Z_2$-invariant scalar potential including the singlet is given by
\begin{eqnarray}
\label{potential}
V & \ = \ &
- \mu^2_L \chi_L^\dagger \chi_L
- \mu^2_R \chi_R^\dagger \chi_R
- \frac12 \mu^2_S S^2  
+ \lambda_1 \left[ (\chi_L^\dagger \chi_L)^2
    +(\chi_R^\dagger \chi_R)^2 \right]
   + \lambda_2 (\chi_L^\dagger \chi_L) (\chi_R^\dagger \chi_R) \nonumber \\
&& + \lambda_S S^4 +  \lambda_3 S^2 (\chi_L^\dagger \chi_L + \chi_R^\dagger \chi_R) \,.
\end{eqnarray}
As in \cite{Mohapatra:2014qva} we keep in the potential the mass terms $\mu^2_{L,\,R}$ different so that they break the LR symmetry softly. Among all the terms given above, the $\lambda_2$ term mixes the left and right-handed doublets and the  $\lambda_3$ term couples the singlet $S$ to the doublets. 
The mass square matrix for the scalars after minimization of the potential is given by
\begin{eqnarray}
{\cal M}^2_{\rm scalar} \ = \ \left( \begin{matrix}
-\mu_L^2 + 3 \lambda_1 v_L^2 & \lambda_2 v_L v_R &  2 \lambda_3 v_L v_S \\
\lambda_2v_L v_R & -\mu_R^2 + 3 \lambda_1 v_R^2 &  2 \lambda_3 v_R v_S \\
2 \lambda_3 v_L v_S & 2 \lambda_3 v_R v_S & - \mu_S^2 + 6 \lambda_S v_S^2
\end{matrix} \right) \, .
\label{eq:mssq}
\end{eqnarray}
The FCNC constraint on the $W_R$ boson from $K$-meson mixing implies a lower bound on the VEV $v_R$ (which depends also on the gauge coupling $g_R$)~\cite{Guadagnoli:2011id}, and is much larger than the SM VEV $v_L$. Then the mixing between the SM Higgs $h$ and its heavy RH partner $H$ is suppressed by the ratio $\lambda_2 v_L /2\lambda_1 v_R$~\cite{Mohapatra:2014qva}. 
The $h - S$ mixing is expected to be of order $\lambda_3 v_L / \lambda_S v_S$. On the contrary, as long as $\lambda_3 \sim \lambda_S$, under the condition of $v_S \sim v_R$ at the (multi-)TeV scale, $H$ and $S$ can potentially have sizable mixing. Then in this case the singlet $S$ can decay into the SM top quark, $W$ and $Z$ gauge bosons and SM Higgs pairs by the large mixing to $H$ \cite{Mohapatra:2014qva}. It is also subject to some constraints from flavor observations~\cite{delAguila:2000rc, Kawasaki:2013apa}, precision tests~\cite{delAguila:2010mx, Aguilar-Saavedra:2013qpa} and neutrinoless double beta decay~\cite{Patra:2012ur}. To explain the 750 GeV diphoton excess, we make the simple choice that  the mixing term  $\lambda_3$ at tree level is very small, so that the SM Higgs observables~\cite{Bonne:2012im, Moreau:2012da, Angelescu:2015kga} and the electroweak vacuum stability analysis~\cite{Mohapatra:2014qva} do not get affected much by the presence of $S$. 

\section{Production and decay of the singlet}

The dominate production channel  for the singlet $S$ at the LHC is the gluon-gluon fusion process  mediated by the TeV-scale vectorlike fermions in the triangle loop. As the heavy fermion couplings  to $S$ are proportional to their masses, it is expected that all three generations of heavy fermions contribute significantly to the gluon fusion loops, which enhances the production cross section by roughly a  factor of $N_f^2$ (with $N_f$ being the number of flavors).
At $\sqrt{s} = 13$ TeV, the cross section for gluon fusion channel for a 750 GeV scalar with SM Higgs-like couplings is $\sigma_0^{\rm 13~TeV} \simeq 850$ fb, taking into account the large $k-$factor of order 2~\cite{Higgs}. The corresponding production cross section of the singlet $S$ in our model can be obtained easily by  rescaling the loop factor $A_{1/2}$ and the Yukawa couplings and masses in the fermion loop: $\sigma(gg\to S)^{\rm 13~TeV}=\sigma_0^{\rm 13~TeV}r$, where the scaling factor is defined as
\begin{eqnarray}
r \ = \ \left| \frac{f_T \sin\alpha_L^t \sin\alpha_R^t}{M_t/v_L}
+ \sum_{F=U,D}  \frac{ f_F v_L}{M_{F}} \frac{ A_{1/2} (\tau_{F})  }{  A_{1/2} (\tau_t) }   \right|^2,
\label{ratio}
\end{eqnarray}
with the first term on the RHS for the top quark loop (with $M_t\simeq 173.2$ GeV) and the second term for all the heavy vectorlike quark partners. Here $\tau_F = M_S^2 / 4 M_F^2$ and the fermion loop function is given by (see e.g.~\cite{Djouadi:2005gi})
\begin{align}
A_{1/2}(\tau) \ = \ 2 \,\big[\tau\,+\,(\tau-1)f(\tau)\big]\, \tau^{-2} \;,
\label{loop1}
\end{align}
where the function $f(\tau)$ is defined as
\begin{align}
f(\tau) \ = \ \left\{ \begin{array}{cc}
{\rm \arcsin}^2\sqrt{\tau} & ({\rm for}~\tau\leq 1) \\
-{\displaystyle \frac{1}{4}}\left[\log \left( \frac{1+\sqrt{1-1/\tau}}{1-\sqrt{1-1/\tau}}\right)-i\pi  \right]^2 & ({\rm for}~\tau>1) \;.
\end{array}\right.
\label{fx}
\end{align}
In the parameter space of interest, the top quark contribution is suppressed by the LH mixing
\begin{eqnarray}
\sin\alpha_L^t \ \simeq \ \frac{y_T v_L}{\sqrt2 f_T v_S} \ \sim \ 0.1 \, ,
\label{tTL}
\end{eqnarray}
where we have chosen $v_S\sim 1.2$ TeV, as is required to explain the diphoton excess (see Figure~\ref{fig:allowed} below) and
 the Yukawa parameter $y_T$ is given by
\begin{eqnarray}
y_T \ = \ \sqrt{\frac{2M_t M_T}{v_L v_R}} \sim {\cal O}(1) \,.
\end{eqnarray}
On the other hand, due to the large Yukawa coupling and $v_R$, the RH $t - T$ mixing is generally very large, of order one, and the top partner mass is largely from or even dominated by the $v_R$ term
\begin{align}
M_T^2 \ = \ \frac12 y_T^2 (v_R^2 + v^2_L) + f_T^2 v_S^2 -M_t^2 .
\label{MT}
\end{align}
For a typical value of $v_R = 3$ TeV and $v_S = 1$ TeV, the top partner mass can reach up to 2.5 TeV.
The LH and RH mixing of other SM quarks and their corresponding heavy partners are comparatively much smaller and can be safely neglected, for which $M_F \simeq f_F v_S$ with $f_F$ a universal Yukawa coupling for all the flavors of vectorlike quarks and leptons in the Lagrangian (\ref{eq:Lyukawa}).\footnote{For the bottom quark partner, the difference $M_B - f_B v_S$ is only 26 GeV for $v_R = 3$ TeV and $f_B v_S=1$ TeV, for instance. For other quark and lepton flavors the mixing effects are even smaller.} The production of $S$ is then predominately from these five heavy quark loops in Eq.~\eqref{ratio}.
For a fixed value of $f_F\sim {\cal O}(1)$, the production cross section is suppressed by the heavy fermion masses via approximately $\sigma (gg \rightarrow S) \propto M_F^{-2}$.

As far as the decay of $S$ is concerned, due to our choice of parameters, it couples dominantly to the heavy vectorlike fermions with a coupling $f_F$ of order one. All the couplings of $S$ to other particles are from mixing or radiative effects. The largest coupling to the SM fermions is to the top quark from $t-T$ mixing, i.e. $f_T \sin\alpha_L^t \sin\alpha_R^t$, which is however suppressed by the small LH mixing $\sin\alpha_L^t$.
For loop-induced channels, $S$ can decays into $gg$, $\gamma\gamma$, $\gamma Z$ and $ZZ$  mainly via the large number of heavy quark and lepton loops.
As the SM $W$ boson does not couple directly to the heavy fermions, the $S \rightarrow WW$ channel is dominantly induced from the top quark loop which is comparatively suppressed by the small mixing $\sin\alpha_L^t$. On the other hand, the $Z$ boson couples to the heavy vectorlike fermions with couplings proportional to $Q_F s_w^2$, where $Q_F$ is the electric charge of heavy fermions and $s_w\equiv \sin\theta_w$, $\theta_w$ being the SM weak mixing angle. As $M_S \gg 2M_Z$, it is expected that the $\gamma\gamma$, $\gamma Z$ and $ZZ$ decay channels have comparable fractional widths
\begin{eqnarray}
\Gamma_{\gamma\gamma} \ : \ \Gamma_{Z\gamma} \ : \ \Gamma_{ZZ} \ = \ 1 \ : \ 2\tan^2\theta_w \ : \  \tan^4\theta_w \,,
\label{brratio}
\end{eqnarray}
which is a clean signal of vectorlike fermions. 
Combining all these decay channels, we find that since they are suppressed either by the small mixing or by the loop factors, the width of hidden singlet $S$ is rather small, as we will show below.

The various partial decay widths of $S$ from tree and loop-level interactions are given by
\begin{align}
  \Gamma_{t\bar{t}}    \ & = \
  \frac{3 y_{S t\bar{t}}^2 M_{S}}{16\pi}
  \left( 1 - \frac{4M_t^2}{M_{S}^2} \right)^{3/2} \,,\label{tt} \\
 \Gamma_{gg}    \ & = \  \frac{\alpha_s^2 M_S^3}{128\pi^3} \left| \sum_{F=t,\,U,\,D}
  \frac{f_F}{M_F} A_{1/2} (\tau_F) \right|^2
  \left( 1 + k_{gg} \frac{\alpha_s}{\pi} \right) \,, \label{gg} \\
 \Gamma_{\gamma\gamma}   \ & = \
  \frac{\alpha^2 M_S^3}{256\pi^3} \left| \sum_{F=t,\,U,\,D,\,E}
  \frac{f_F}{M_F} N_{C_F} Q_F^2 A_{1/2} (\tau_F)
  \right|^2 \,,\label{gamma} \\
 \Gamma_{\gamma Z}   \ & \simeq  \
  \frac{\alpha^2 M_S^3}{128\pi^3 s_w^2 c_w^2}  \left| \sum_{F=t,\,U,\,D,\,E}
  \frac{f_F}{M_F} N_{C_F} Q_F \left( \frac12 I_{3_F} - Q_F s_w^2 \right) A_{1/2} (\tau_F)
  \right|^2 \,, \label{Z}\\
 \Gamma_{ZZ}   \ & \simeq  \
  \frac{\alpha^2 M_S^3}{256\pi^3 s_w^4 c_w^4}  \left| \sum_{F=t,\,U,\,D,\,E}
  \frac{f_F}{M_F} N_{C_F} \left( \frac12 I_{3_F} - Q_F s_w^2 \right)^2 A_{1/2} (\tau_F)
  \right|^2 \,. \label{ZZ}
\end{align}
For the $\gamma Z$ and $ZZ$ channels we show the formulae in the massless $Z$ boson limit in which we can see clearly the relation given in Eq.~\eqref{brratio}. The exact expressions for these two channels are given in the Appendix. In Eqs.~\eqref{tt}-\eqref{ZZ}, $\tau_F = M_S^2 / 4 M_F^2$,  $y_{St\bar{t}} = \sqrt2 f_T  \sin\alpha_L^t \sin\alpha_R^t$ and $k_{gg} = \frac{95}{4} - \frac{7N_f}{6}$ (with $N_f =6$) is the next-to-leading order factor for the gluon decay channel. The strong coupling $\alpha_s\equiv g_s^2/4\pi$ and the fine-structure constant $\alpha\equiv e^2/4\pi$ are evaluated at the resonance scale of 750 GeV. The loops run over the top and heavy vectorlike fermions for which the effective Yukawa couplings $f_F$ are respectively $f_T \sin\alpha_L^t \sin\alpha_R^t$ (top quark), $f_T \cos\alpha_R^t$ (top partner) and $f_F$ (all other heavy fermions).

\begin{figure}[t!]
  \centering
  \includegraphics[width=0.5\textwidth]{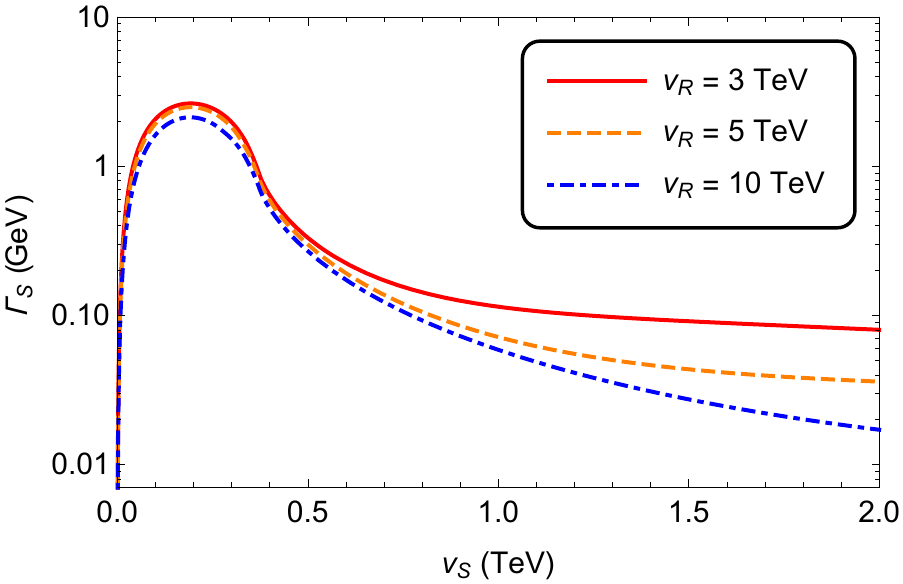}
  \caption{Total width of the scalar $S$ in the LR seesaw model, as a function of the singlet VEV $v_S$. The different curves are for $v_R = 3$, 5 and 10 TeV. Here we have set $f_F=1$. }
  \label{fig:width}
\end{figure}

The total decay width of the scalar $S$ is then given by
\begin{align}
\Gamma_S  \ \simeq \  \Gamma_{t\bar{t}}+\Gamma_{gg}+\Gamma_{\gamma\gamma}+\Gamma_{\gamma Z}+\Gamma_{ZZ}.
\label{tot}
\end{align}
In the minimal LR seesaw model, only the universal Yukawa coupling $f_F$ and the VEV $v_S$ are input by hand, apart from the RH scale $v_R$. The total decay width for some typical values of $v_R$ is shown in Figure~\ref{fig:width} as a function of the VEV $v_S$. For this and subsequent plots, we have set the universal Yukawa coupling $f_F = 1$, unless otherwise specified.
 Note that the total width of $S$ is at most a few GeV and decreases with higher $v_S$. Therefore, if this model is right, $S$ will always appear as a narrow resonance at the LHC. This prediction can be easily tested in near future. In particular, if the wide resonance behavior of the reported excess with $\Gamma_S\sim 45$ GeV as claimed by ATLAS~\cite{ATLAS} persists with more data, the minimal version of this model will be ruled out.

\begin{figure}[t!]
  \centering
  \includegraphics[width=0.48\textwidth]{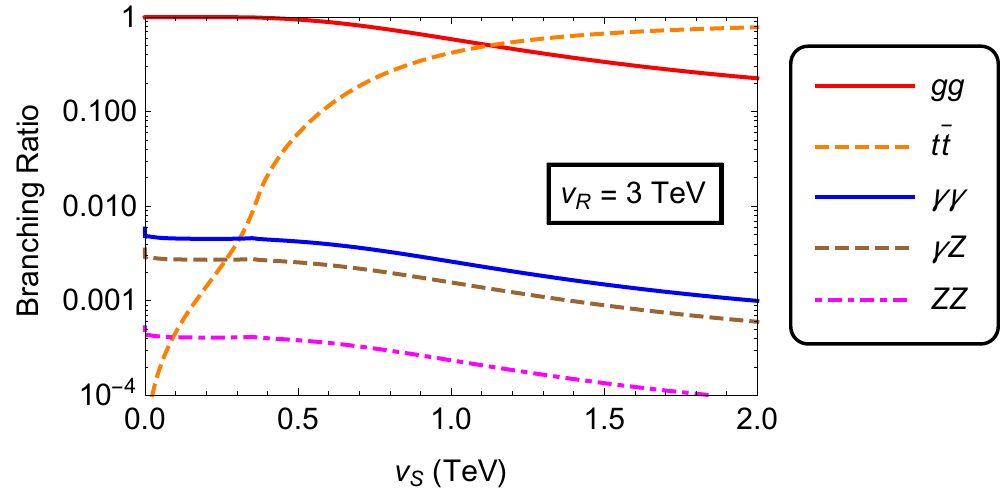}
  \includegraphics[width=0.48\textwidth]{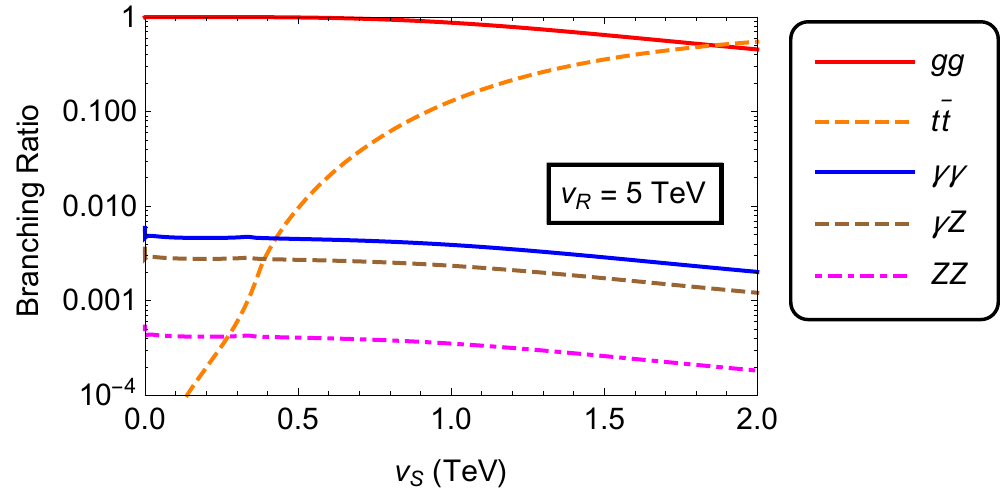}
  \caption{Branching ratios of the scalar $S$ into $t\bar{t}$, $gg$, $\gamma\gamma$, $\gamma Z$ and $ZZ$ final states in the LR seesaw model, as functions of the singlet VEV $v_S$. The two panels are for $v_R = 3$ and 5 TeV respectively.}
  \label{fig:br}
\end{figure}

The various branching ratios for the channels discussed above are shown in Figure~\ref{fig:br} as a function of the VEV $v_S$ for $v_R = 3$ and 5 TeV. Comparing the different decay channels, we find that in most of the parameter space of interest, $gg$ is the dominate decay channel. The $t\bar{t}$ channel depends quadratically on the Yukawa coupling and increases when $v_S$ is larger, which can be easily seen from the seesaw relation (\ref{seesaw}). For $v_S \sim 1$ TeV it is even comparable with the loop-induced $gg$ channel. The decay rates into SM gauge bosons $\gamma\gamma$, $\gamma Z$ and $ZZ$ are suppressed by both the gauge coupling $\alpha^2 / \alpha_s^2$ and loop factors and are at the per mil level or even smaller.

\begin{figure}[t!]
  \centering
  \includegraphics[width=0.6\textwidth]{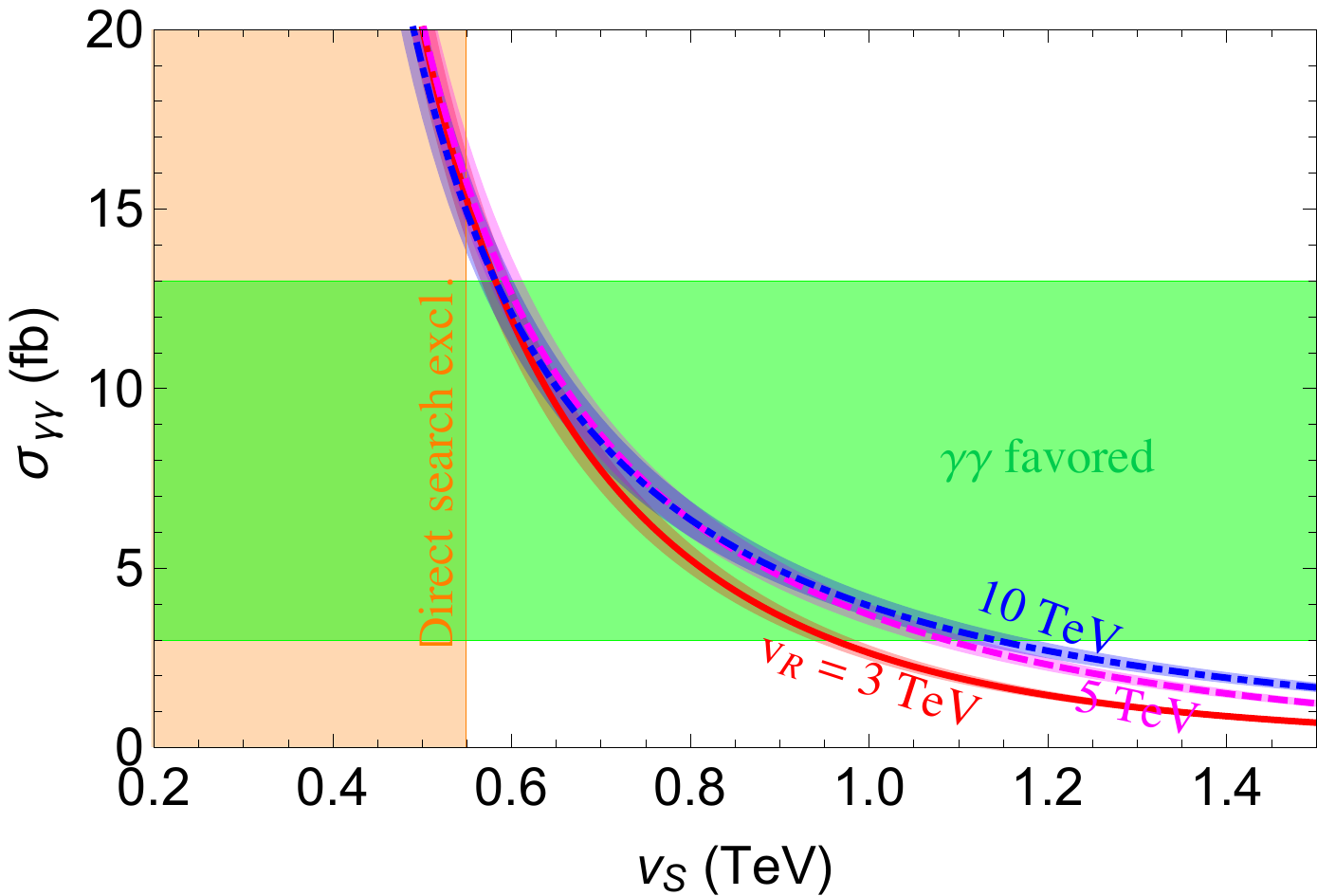}
  \caption{The LR seesaw parameter space fitting the $\sqrt s=13$ TeV diphoton excess, as a function of the singlet VEV $v_S$. The bands around each line for a given $v_R$ shows the uncertainties from parton distribution functions and $\alpha_s$~\cite{error}. The horizontal (green) band shows the best-fit range to explain the diphoton excess at the $\sqrt s=13$ TeV LHC~\cite{CMS, ATLAS}, whereas the vertical (orange) shaded region is ruled out from direct searches for heavy vectorlike quarks at $\sqrt s=8$ TeV LHC~\cite{Aad:2015kqa}. }
  \label{fig:cs}
\end{figure}

The diphoton signal cross section $\sigma_{\gamma\gamma}\equiv \sigma(gg\rightarrow S)\times {\rm BR}({S\to \gamma\gamma})$ as function of  $v_S$ is shown in Fig.~\ref{fig:cs} for different values of the RH scale $v_R = 3$, 5 and 10 TeV.
The horizontal (green) shaded region shows the preferred range of the observed diphoton signal: $\sigma_{\gamma\gamma}^{\rm obs}=(8\pm 5)$ fb~\cite{CMS, ATLAS}.  The vertical (orange) shaded region is the 95\% CL exclusion region from direct searches for the bottom partner at the $\sqrt s=8$ TeV LHC~\cite{Aad:2015kqa}.  The corresponding limits for the top partner mass are always satisfied for the choice of $v_R$ in our model. We note that the signal cross section decreases when the vectorlike fermion mass is heavier, as expected. Furthermore, the mild dependence on the RH scale $v_R$ in some region of the parameter space is mainly from the mixing of SM quarks and their heavy partners. 


In order to see whether the allowed parameter space shown in Fig.~\ref{fig:cs} is consistent with the upper limits on signal cross sections in other channels from the publicly available $\sqrt s=8$ TeV data, we evaluate the coresponding production cross section $\sigma(gg\to S)^{\rm 8~TeV}=\sigma_0^{\rm 8~TeV}r$, where $\sigma_0^{\rm 8~TeV}=156.8$ fb~\cite{Higgs} and $r$ is the scaling factor defined in Eq.~\eqref{ratio}.  Using Eqs.~\eqref{tt}-\eqref{ZZ} and \eqref{tot}, we calculate the corresponding signal cross sections $\sigma_{XY}^{\rm 8~TeV}\equiv \sigma(gg\to S)^{\rm 8~TeV}\times {\rm BR}(S\to XY)$, where $XY=t\bar{t}, gg, \gamma\gamma, \gamma Z$ and $ZZ$, respectively. These values are to be compared with the 95\% CL upper limits on the cross sections obtained with the $\sqrt s=8$ TeV LHC data~\cite{Khachatryan:2015sma, Aad:2014aqa, CMS:2015neg, CMS:2014onr, Aad:2015mna, Aad:2014fha, Aad:2015kna}, namely,
\begin{eqnarray}
\sigma_{t\bar{t}} \  < \ 450~{\rm fb} , \quad 
\sigma_{gg} \ < \  2.5~{\rm pb}, \quad 
\sigma_{\gamma\gamma} \ < \ 1.5~{\rm fb}, \quad 
\sigma_{\gamma Z}\ < \ 4~{\rm fb}, \quad 
\sigma_{ZZ}\ < \ 12~{\rm fb}. \label{ZZ8}
\end{eqnarray}
In Fig.~\ref{fig:allowed}, we compare the parameter space in the $(v_S,\, v_R)$ plane as preferred by the $\sqrt s=13$ TeV diphoton excess (green shaded region) with the $\sqrt s=8$ TeV exclusion limits given by Eq.~\eqref{ZZ8}. We find that the most stringent constraint comes from the $\sqrt s=8$ TeV diphoton search, as shown by the red shaded region in Fig.~\ref{fig:allowed}, which disfavors part of the $\sqrt s=13$ TeV diphoton excess region. The $\gamma Z$ and dijet constraints also rule out the low vectorlike fermion mass region, as shown by the magenta and blue shaded regions, respectively. The $t\bar{t}$ and $ZZ$ limits are much weaker in this model and only rule out the very low $v_R$ and $M_F$ values (not shown in this plot).
The direct search limit~\cite{Aad:2015kqa} rules out the parameter space with $M_B<575$ GeV, as shown by the orange shaded region. In any case, we find that $M_F$ ($F\neq T$) values between about 700--1150 GeV are still compatible with the existing constraints and can explain the observed diphoton excess in this model.
\begin{figure}[t!]
  \centering
  \includegraphics[width=0.5\textwidth]{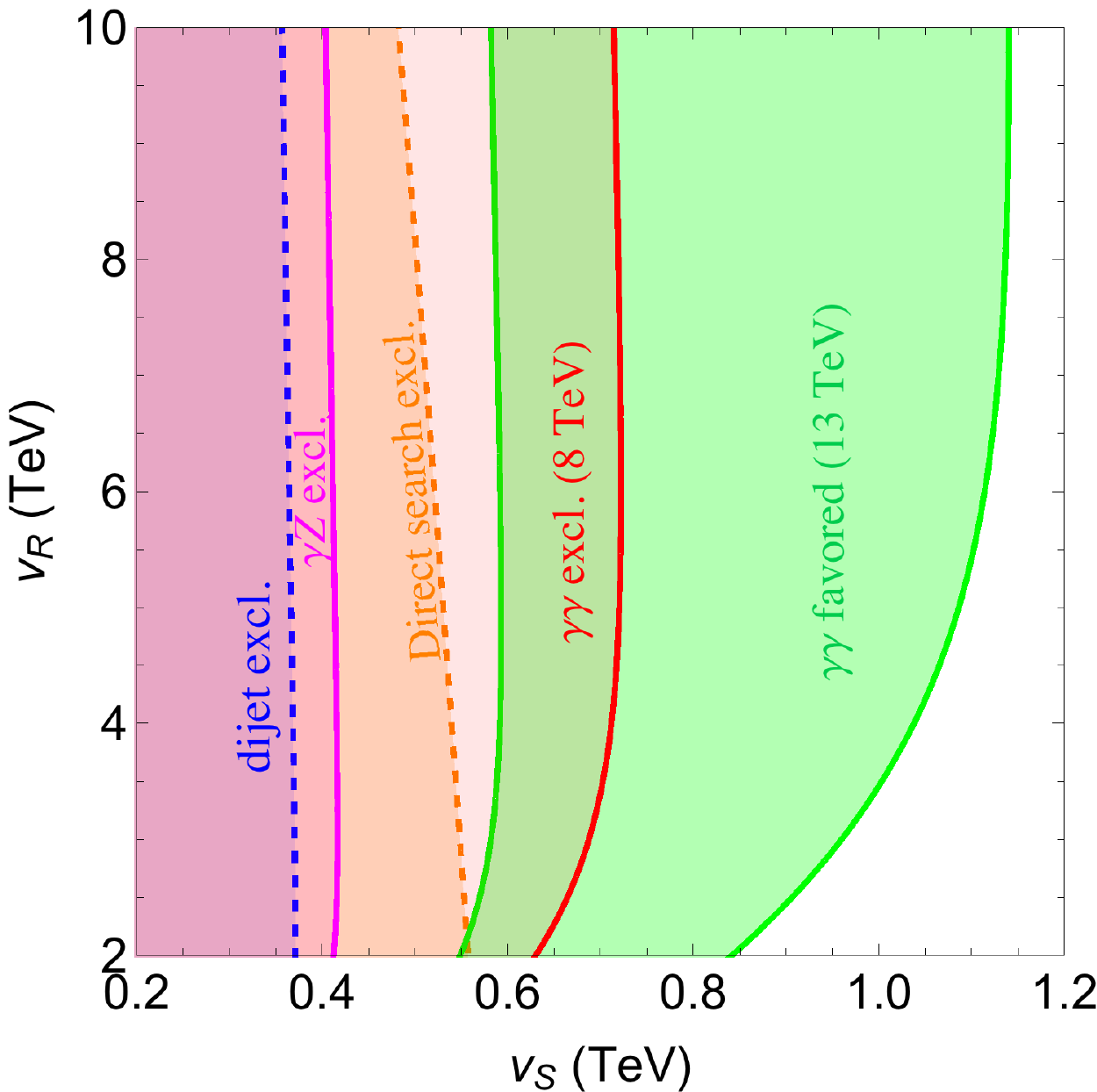}
  \caption{The LR seesaw parameter space fitting the $\sqrt s=13$ TeV diphoton excess (green), compared with the various  exclusion limits from $\sqrt s=8$ TeV LHC data.  }
  \label{fig:allowed}
\end{figure}

\begin{figure}[t!]
  \centering
  \includegraphics[width=0.5\textwidth]{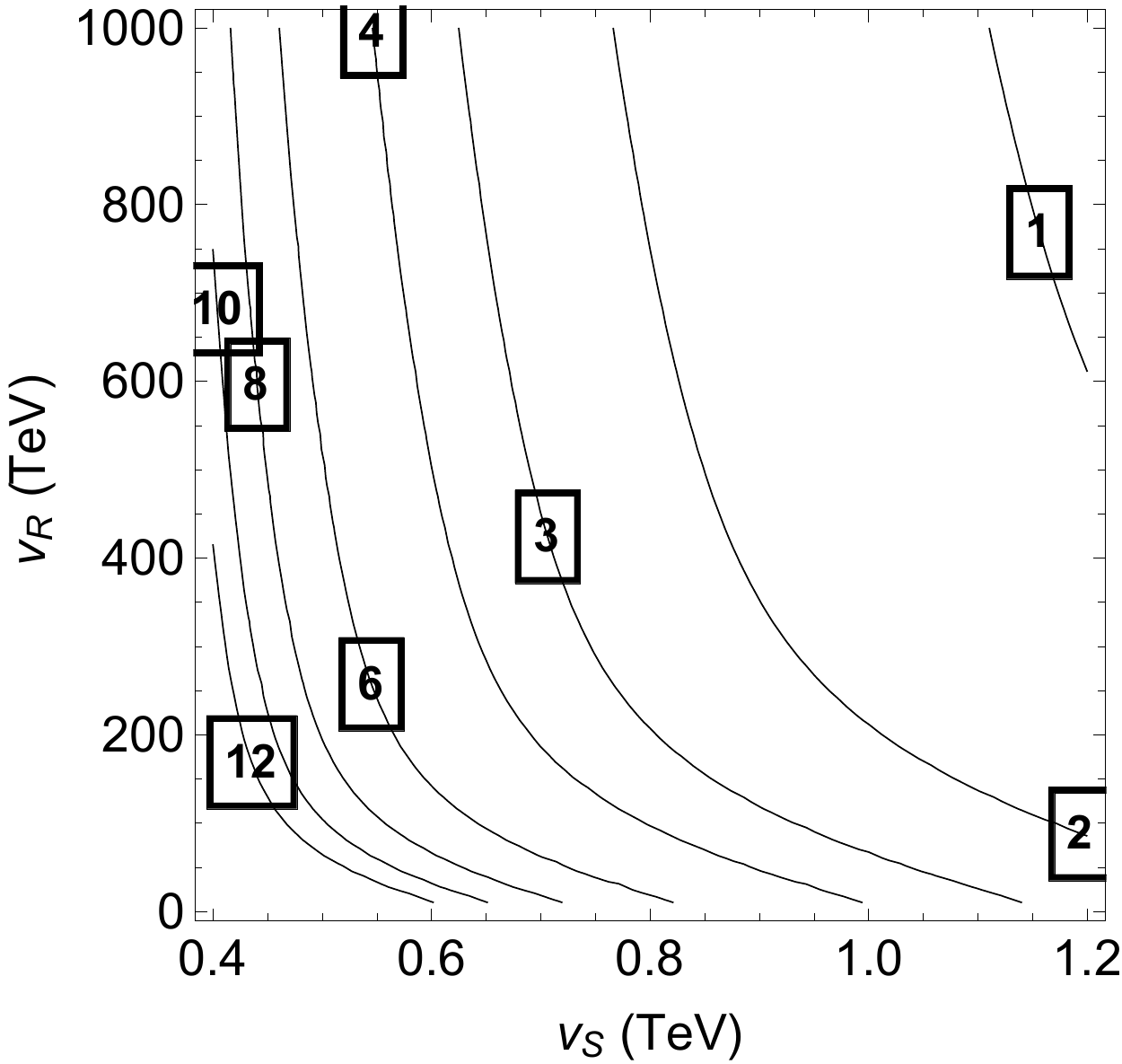}
  \caption{Contours of fixed $\sigma_{\gamma\gamma}$ (in fb) in the $(v_S,\, v_R)$ plane.  } 
  \label{fig:upper}
\end{figure}

It is interesting to note that for a fixed value of the singlet VEV $v_S$, there exists an {\it upper} limit on the RH scale $v_R$. This is shown in Figure~\ref{fig:upper} where the different contours show the diphoton signal cross section $\sigma_{\gamma\gamma}$ in fb at $\sqrt s=13$ TeV LHC. We find that for $v_S=1$ TeV, $v_R$ must be smaller than 65 TeV to satisfy the observed value of $\sigma_{\gamma\gamma}$. For higher values of $v_S$, the upper limit on $v_R$ becomes stronger. This provides another potential way to distinguish this model from other explanations of the diphoton excess. 

For the testability of this model in the ongoing run of the $\sqrt s=13$ TeV LHC in near future, we choose two optimum sets of parameters with the RH scale $v_R = 3$ TeV and 5 TeV respectively, for which the total production cross section, total decay width, the dominant fractional width and branching ratios of $S$ are listed in Table~\ref{tab:example}.

\begin{table}[t]
  \centering
  \caption[]{Input parameters ($f_F$ and $v_S$) for $v_R$ = 3 TeV and 5 TeV, the total production cross section and decay width of the scalar $S$, and the signal cross sections in various relevant channels at the $\sqrt s=13$ TeV LHC in the minimal LR seesaw model.  }
  \label{tab:example}
  \begin{tabular}{|c|c|c|c|c|}
  \hline\hline
   & \multicolumn{2}{c|}{$v_R$ = 3 TeV} &  \multicolumn{2}{c|}{$v_R$ = 5 TeV} \\ \hline\hline
  $f_F$ (input) & \multicolumn{2}{c|}{1} & \multicolumn{2}{c|} {1} \\ \hline
  $v_S$ [GeV] (input) & \multicolumn{2}{c|}{800} & \multicolumn{2}{c|}{1000} \\ \hline
  $\sigma_{} (gg\rightarrow S)$ [pb] & \multicolumn{2}{c|}{1.61} & \multicolumn{2}{c|}{0.95} \\ \hline
  $\Gamma_{\rm total} (S)$ [GeV] & \multicolumn{2}{c|}{0.21} & \multicolumn{2}{c|}{0.071} \\ \hline\hline
  \multicolumn{5}{|c|}{ signal cross section [fb] } \\ \hline
  $t\bar{t}$ & \multicolumn{2}{c|}{423} & \multicolumn{2}{c|}{122} \\ \hline
  $gg$ & \multicolumn{2}{c|}{1173} & \multicolumn{2}{c|}{825} \\ \hline
  $\gamma\gamma$ & \multicolumn{2}{c|}{5.3} & \multicolumn{2}{c|}{3.7} \\ \hline
  $\gamma Z$ & \multicolumn{2}{c|}{3.2} & \multicolumn{2}{c|}{2.3} \\ \hline
  $ZZ$ & \multicolumn{2}{c|}{0.48} & \multicolumn{2}{c|}{0.34} \\ \hline
  \end{tabular}
  \end{table}

\section{High-Scale Validity} \label{sec:rge}
\begin{figure}[t!]
\centering
\includegraphics[width=7cm]{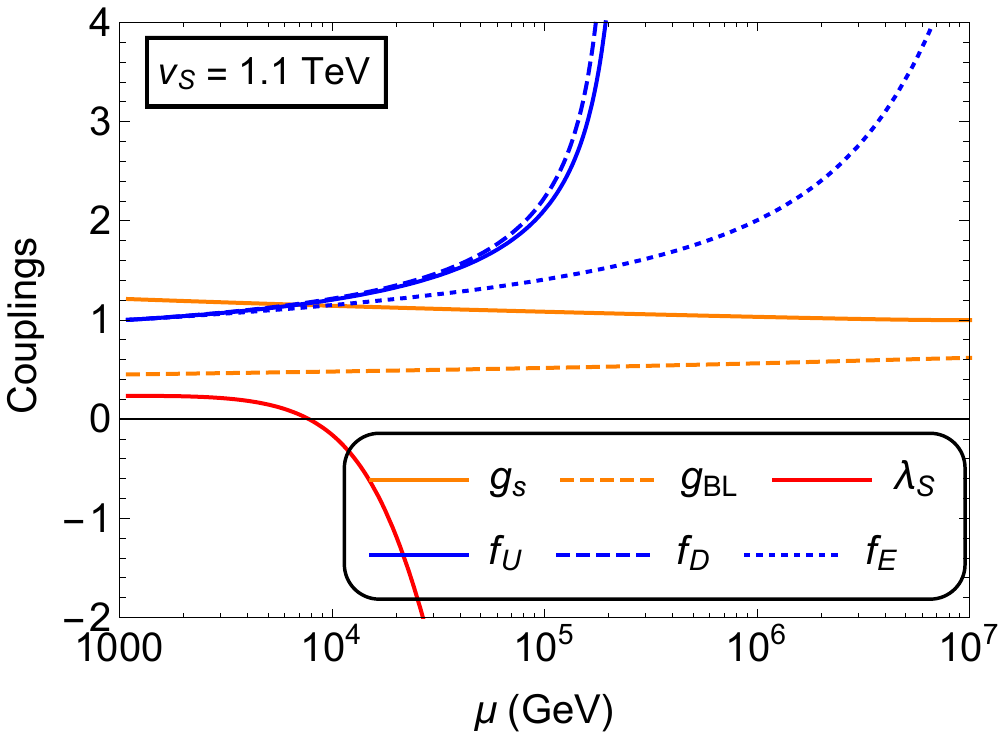}
\hspace{0.5cm}
\includegraphics[width=7cm]{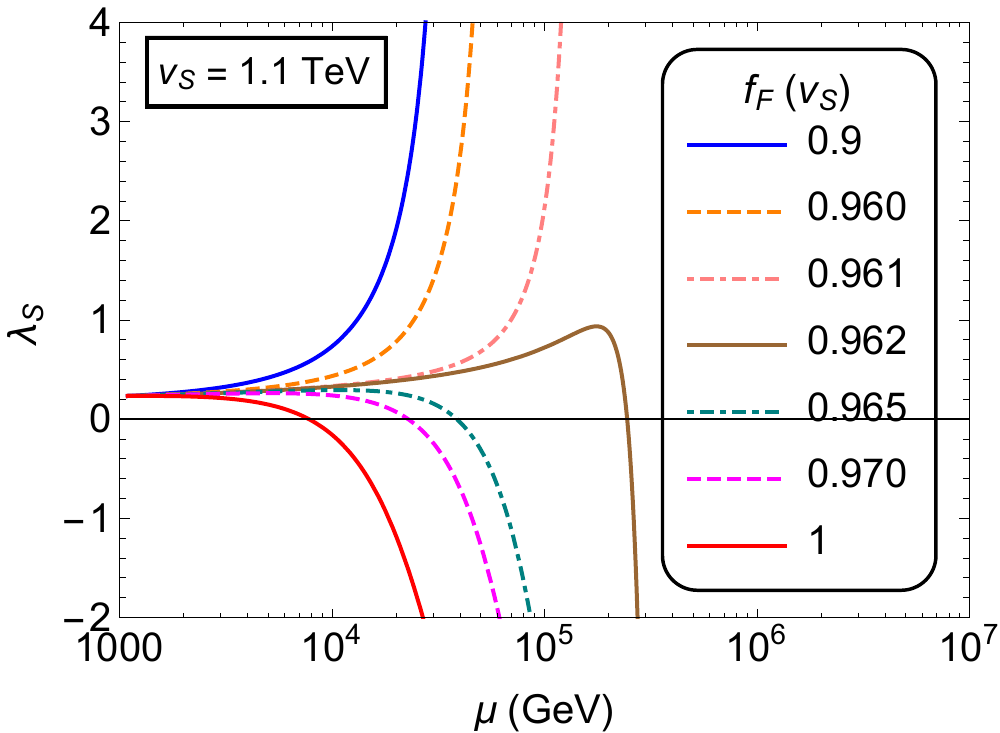}
\caption{(Left) The one-loop RG evolution of the couplings given by Eqs.~\eqref{eq:rg1}-\eqref{eq:rg2} for a typical choice of the singlet VEV $v_S=1.1$ TeV and with the initial condition  $f_U=f_D=f_E=0.98$ at $\mu=v_S$. (Right) The dependence of the $\lambda_S$ running on the initial conditions for $f_{U,D,E} (v_S)$.}
\label{fig:rg}
\end{figure}

Due to the presence of a large number of extra fermions in our model and their ${\cal O}(1)$ Yukawa couplings with the singlet scalar field $S$, it is expected that the singlet scalar self-coupling $\lambda_S$ would become negative well before the Planck scale, thus potentially destabilizing the singlet vacuum. This would mean that our TeV-scale model provides only an effective description of the 750 GeV diphoton excess and must be augmented by some other new physics for its ultraviolet (UV)-completion. In order to find this cut-off scale,  we write down the one-loop renormalization group (RG) equation for the scalar self-coupling $\lambda_S$, as well as for other relevant couplings affecting the evolution of $\lambda_S$  above the scale $v_S$:\footnote{See also Refs.~\cite{Dhuria:2015ufo, Salvio:2015jgu, Hamada:2015skp} for similar RG analyses in presence of vectorlike fermions.} 
\begin{align}
16\pi^2\frac{dg_s}{dt} & \ = \ -3g_s^3 \, , \label{eq:rg1} \\
16\pi^2\frac{dg_{BL}}{dt} & \ = \ \frac{41}{2}g_{BL}^3 \, , \\ 
16\pi^2\frac{d\lambda_S}{dt} & \ = \ 72\lambda_S^2 +24\lambda_S\left[3(f_U^2+f_D^2)+f_E^2\right]-6\left[3(f_U^4+f_D^4)+f_E^4\right], \label{eq:rg3} \\
16\pi^2\frac{df_U}{dt} & \ = \ 21f_U^3-8g_s^2f_U-\frac{8}{3}g_{BL}^2f_U\, ,\\
16\pi^2\frac{df_D}{dt} & \ = \ 21f_D^3-8g_s^2f_D-\frac{2}{3}g_{BL}^2f_D\, ,\\
16\pi^2\frac{df_E}{dt} & \ = \ 9f_E^3-2g_{BL}^2f_E\, , \label{eq:rg2}
\end{align}
where $t\equiv \log(\mu)$ is the renormalization scale, and for simplicity, we have assumed a common  Yukawa coupling $f_F$ (with $F=U,D,E$) for each of the three generations of the vectorlike fermions. We have further assumed the coupling $\lambda_3$ in Eq.~\eqref{potential} to be very small, so that it does not affect the RG evolution of other couplings and the electroweak vacuum stability analysis, as given in Ref.~\cite{Mohapatra:2014qva}. For a given value of the singlet VEV $v_S$, the initial value of $\lambda_S(v_S)$ is fixed by Eq.~\eqref{eq:mssq} which, for small $\lambda_3$, reads $\lambda_S (v_S) \simeq \frac{m_S^2}{2v_S^2}$ with $m_S=750$ GeV. For illustration, we choose $v_S=1.1$ TeV which is close to the maximum value allowed by the diphoton excess [cf. Figure~\ref{fig:allowed}]. 
For the vectorlike fermion couplings $f_F(v_S)=1$, we find a cut-off scale around $10$ TeV, as shown in Figure~\ref{fig:rg} (left). For larger initial values of the Yukawa couplings $f_F$, the cut-off scale will be even lower, as is evident from Eq.~\eqref{eq:rg3}. For slightly smaller values of $f_F$, the cut-off scale can be pushed up to about 300 TeV, as can be seen from Figure~\ref{fig:rg} (right). But as we further lower the initial value of $f_F$, the positive contribution to the RHS of Eq.~\eqref{eq:rg3} overtakes the negative one, thus resulting in non-perturbative values of  $\lambda_S$. In any case, the diphoton signal rate also decreases with $f_F$ [cf.~Eq.~\eqref{ratio}], while the direct search constraints on $M_F\simeq f_F v_S$  require a larger $v_S$. Thus, it turns out that we must have $f_F(v_S)\gtrsim 0.5$ to allow some model parameter space satisfying the diphoton excess. 
 
\section{Discussions}

In general vectorlike fermion and (pseudo)scalar interpretations of the diphoton events recently observed at 750 GeV, the Yukawa couplings of vectorlike quarks to the scalar resonance tend to be very large, or the electric charges of the heavy fermions are exotically large. It is preferable in the LR seesaw model that these problems are avoided by the naturally large number of heavy partners for the seesaw generation of SM fermion masses. We emphasize here that while the main features of the diphoton explanation in this model are quite similar to earlier works which introduced a SM-singlet scalar and vectorlike fermions, the specific nature of our model constrained by the strong $CP$ requirement, i.e. seesaw quark masses, leads to some differences. In particular, our minimal LR interpretation of the diphoton excess implies the following features which could be tested by the upcoming LHC data and future precision measurements:
\begin{itemize}
  \item [(i)] As the gluon fusion production for $S$ goes down like $M_F^{-2}$, the observation of diphoton events with cross section at the level of few fb implies the existence of heavy fermions at the TeV scale.

  \item [(ii)] 
      In the minimal LR seesaw model, for heavy partners at the scale of 1 TeV, the RH scale lies naturally at the few TeV scale. In such scenarios we expect a heavy $W_R$ nearby which
  should be accessible at the $\sqrt s = 13$ TeV LHC.\footnote{The recent diboson excess~\cite{Aad:2015owa, Aad:2015ipg} suggests a $W_R$ with mass around 1.9--2 TeV and $g_R=$0.4--0.5 (see e.g.~\cite{Dobrescu:2015qna,Gao:2015irw, Brehmer:2015cia,Dev:2015pga, Deppisch:2015cua, Dias:2015mhm, Brehmer:2015dan}). This scenario still remains to be tested by the $\sqrt s=13$ TeV data. }

  \item [(iii)] The decay of $S$ in the LR model is suppressed either by small mixing parameters or the loop factors, and thus it would be a narrow resonance and could be easily tested by the upcoming LHC data.

  \item [(iv)] A large portion of the $S$ decay width is into top quark pairs, with a branching ratio varying from a few percent to half or even larger (subject to the top quark mixing and other parameters). Thus, the $t\bar{t}$ observations could not only test the minimal LR model but also be used to probe the $t-T$ mixing which is essential to search for new physics in the top quark sector.

  \item [(v)] In the minimal scenario of the LR seesaw model, the singlet $S$ decouples from the doublets for gauge symmetry breaking. In such cases the stability analysis of the scalar potential is the same as in \cite{Mohapatra:2014qva}. In a large parameter space the SM vacuum is stable up to the GUT scale or even the Planck scale.

\item [(vi)] The presence of the new scalar $S$ and the vectorlike fermions has a potential impact on the precision electroweak observables~\cite{Xiao:2014kba}. However, with the small $t-T$ mixing angle $\sin\alpha_L^t$ as in our case [cf.~Eq.~\eqref{tTL}], these constraints are easily satisfied for the range of top-partner masses considered here. In future, a better measurement of the top-quark decay width which has currently a large uncertainty of ${\cal O}$(1 GeV)~\cite{Abazov:2012vd, Aaltonen:2013kna} could be another way to test our hypothesis.  

\item [(vii)] Due to the presence of a large number of vectorlike fermions in the model, the scalar self-coupling becomes negative at around 10-100 TeV (see Section~\ref{sec:rge}), thus signaling the onset of some other new physics for its UV-completion. This new physics could be connected with the appearance of an expanded gauge group, e.g. $SU(2)_L\times SU(2)_R\times U(1)_{B-L, L}\times U(1)_{B-L,R}$ or $SU(5)\times SU(5)$. These topics are currently under investigation. 
\end{itemize}

\section{Conclusion}
We have presented a left-right quark seesaw model with TeV-scale vectorlike fermions and a singlet scalar, where the singlet scalar responsible for the mass of the vectorlike fermions can be identified as the new 750 GeV resonance indicated by the early $\sqrt s=13$ TeV LHC data.  This model was originally proposed to solve the strong CP problem without the axion. We have discussed various ways this hypothesis can be tested in near future as more data is amassed in the run II of the LHC.


\section*{Acknowledgments}

Y.Z. thanks Julian Heeck for inspiring discussions on the singlet in LR seesaw model. The work of P.S.B.D is supported by the DFG with grant RO 2516/5-1. The work of R.N.M. is supported in part by the US National Science Foundation Grant No. PHY-1315155. Y.Z. would like to thank the IISN and Belgian Science Policy (IAP VII/37) for support.

\appendix
\section{Exact formulae for the $\gamma Z$ and $ZZ$ channel}\label{gammaZ}

The exact formulae for the loop-induced $\gamma Z$ and $ZZ$ decay channels of $S$ are respectively
\begin{align}
 \Gamma_{\gamma Z}    \ & =
\   \frac{  \alpha^2 M_S^3}{32\pi^3s_w^2c_w^2} \left( 1 - \frac{M_Z^2}{M_S^2} \right)^3
  \left| \sum_{i=t,\,P,\, N,\,E}
  \frac{f_i }{M_i} N_{Ci} Q_i \left( I_{3i} - 2Q_i s_w^2 \right) A_{1/2} (\tau_i^{-1},\,\lambda_i^{-1})
  \right|^2 \,, \label{Zexact}\\
 \Gamma_{ZZ}   \ & = \
  \frac{\alpha^2 M_S^3}{256\pi^3 s_w^4 c_w^4} \left( 1 - \frac{4M_Z^2}{M_S^2} \right)^{1/2} \left| \sum_{i=t,\,P,\,N,\,E}
  \frac{f_i}{M_i} N_{Ci} \left( \frac12 I_{3i} - Q_i s_w^2 \right)^2 A_{ZZ} (\tau_i, \, \lambda_i)
  \right|^2 \,. \label{ZZexact}
\end{align}
The loop function  $A_{1/2} (\tau,\,\lambda)$ is given by~\cite{Djouadi:2005gi}
\begin{eqnarray}
A_{1/2}(\tau, \lambda)  \ & = & \   \frac{\tau \lambda}{2(\tau-\lambda)} +\frac{\tau^2 \lambda^2}{2(\tau - \lambda)^2}\,\left[f(\tau^{-1}) - f(\lambda^{-1})\right]
  +  \frac{\tau^2 \lambda}{(\tau - \lambda)^2}\, \left[(g(\tau^{-1}) - g(\lambda^{-1})\right] \nonumber \\
&& +\frac{\tau\lambda}{2(\tau-\lambda}\left[f(\tau^{-1}) - f(\lambda^{-1}) \right],
\label{loop2}
\end{eqnarray}
where the function $f(x)$ is defined in Eq.~\eqref{fx} and $g(x)$ is defined as
\begin{equation}
g(x) =
\left\{
    \begin{array}{ll}
                \frac{\sqrt{1 - 1/x}}{2}\left[\log\left(\frac{1 + \sqrt{1 - 1/x}}{1 - \sqrt{1 - 1/x}}\right) - i\pi\right]& \mbox{if } x < 1 \\
                \sqrt{1/x-1}\: \arcsin(\sqrt{x})  & \mbox{if } x \geq 1 \, .
    \end{array}
\right.
\end{equation}
For the $ZZ$ channel,
\begin{eqnarray}
A^2_{ZZ} (\tau,\,\lambda) & \ = \ & \frac{2}{\tau^2}
 \left[ 2 + 2 |f_1 | \left( \frac{1-\tau}{\tau} + \frac{1-\beta^4}{4\beta^2} \right) - |f_2| \frac{ (1-\beta^2) }{\beta^2} \right. \nonumber \\
&& +  |f_1 |^2 \Big\{ 3 \tau ^2 -8 \beta ^2 (\tau
   -1) \tau  +2 \beta ^4 \left(5 \tau ^2-16 \tau +8\right) +8 \beta ^6 (\tau -1) \tau + 3 \beta ^8 \tau ^2  \Big\}  \nonumber \\
&& \left. +  |f_2|^2 \frac{3 \left(1-\beta ^2\right)^2}{8 \beta ^4} +
\frac{|f_1 f_2|}{8\beta^4 \tau}
\left(1-\beta ^2\right) \left( -3 \tau +4 \beta ^2 (\tau -1) +3 \beta ^4 \tau \right)  \right] \,,
\end{eqnarray}
where $\tau = M_S^2/4M_F^2$, $\lambda = M_Z^2/4M_F^2$, $\beta = \sqrt{1-4M_Z^2/M_S^2}$ and
\begin{eqnarray}
f_1 (\tau,\lambda) &\ = \ & - \int_{0}^{1} dx \int_{0}^{1-x} dy \frac{4\tau}
{1 - 4\left[ x(1-x) + y(1-y) \right] \lambda -4xy (\tau - 2\lambda) } \,, \\
f_2 (\tau,\,\lambda) & \ = \ &  2f_3 (\tau) - 2f_3(\lambda) \,, \\ 
f_3(\tau) & \ = \ &\left\{ \begin{array}{ll}
-2 + 2 \sqrt{\frac{1}{\tau }-1} \arcsin
   \left(\sqrt{\tau }\right) & ({\rm for}~\tau\leq 1) \\
-2 + i\pi \sqrt{1-1/\tau} + \sqrt{1-1/\tau} \tanh ^{-1}\left(\frac{2 \sqrt{(\tau -1) \tau }}{2
   \tau -1}\right) & ({\rm for}~\tau>1) \;.
\end{array}\right.
\end{eqnarray}
In the limit of $\beta \rightarrow 1$, only the function $f_1(\tau)$ contributes and the loop function $A_{ZZ} (\tau,\,\lambda) \rightarrow A_{1/2} (\tau)$, thus yielding Eqs.~\eqref{Z} and \eqref{ZZ}.

\end{document}